\documentclass[twocolumn,showpacs,preprintnumbers,amsmath,amssymb,aps]{revtex4}
\usepackage{graphicx}
\usepackage{dcolumn}

\usepackage{color}

\newcommand{\be}{\begin{equation}}
\newcommand{\ee}{\end{equation}}
\newcommand{\bea}{\begin{eqnarray}}
\newcommand{\eea}{\end{eqnarray}}

\newcommand{\kk}{{\bf k}}
\newcommand{\tr}{\mbox{tr}}
\newcommand{\zzeta}{{\boldsymbol\zeta}}

\begin{document}
\title{Description of modulated beam dynamics}
\author{Nikolai A. Yampolsky}
\affiliation{Los Alamos National Laboratory, Los Alamos, New Mexico, 87545, USA}
\date{\today}

\begin{abstract}
General formalism for describing dynamics of modulated beams along linear beamlines is developed. We describe modulated beams with spectral distribution function which represents Fourier transform of
the conventional beam distribution function in the 6-dimensional phase space.
The introduced spectral distribution function is localized in some region of the spectral domain for nearly monochromatic modulations
and it can be characterized with a small number of typical parameters which we choose to be the lowest order moments of the spectral distribution. We study evolution of modulated beams in linear beamlines
and find that characteristic spectral parameters transform linearly. The developed approach significantly simplifies analysis of various schemes proposed for seeding X-ray free electron lasers. We use
this approach to study several recently proposed schemes and find the bandwidth of the output bunching in each case.

\end{abstract}

\maketitle

\section{Introduction}
Free electron lasers (FELs) are the only currently available sources capable of generating coherent X-ray radiation. 
Presently several hard X-ray FELs (XFELs) are operated or under construction \cite{FLASH,LCLS,Spring8,EuroXFEL,FERMI}. 
These light sources are designed to work in the Self-Amplified Spontaneous Emission (SASE) regime \cite{SASE} which amplifies shot noise within the FEL bandwidth.
As a result, the output X-ray pulses have limited longitudinal coherence which may be undesirable for some applications. The XFEL performance can be strongly improved if it is
seeded with coherent signal. Seeded FELs generate more powerful radiation with smaller bandwidth and their undulators are shorter.
That allows to design cheaper light sources having significantly larger brightness.

Growing FEL mode couples radiation with electron motion. Therefore, FEL can be seeded either with narrowband radiation \cite{optical_seeding} or with electron beam
modulated at the X-ray wavelength \cite{HGHG,EEHG,CHG}. The second option looks attractive since electrons are charged particles and they interact with electromagnetic fields unlike radiation which
weakly interacts with materials at X-ray frequencies. Large variety of available beam optics elements introduces a lot of options for transforming modulation and controlling parameters of the output
beam. A number of different schemes for generating modulated beams were recently proposed \cite{HGHG,CHG,EEHG,EEHG-EEX,JMO} and some were verified experimentally \cite{HGHG_exp,EEX_exp,EEHG_exp}.
All these schemes utilize the same principle: first the beam is modulated and then the beam modulation is transformed in the following beamline to produce bunched beam.
Conventional analysis of each scheme relies on following electron trajectory
in the phase space. Such an analysis is somewhat complicated and does not provide intuitive understanding of the scheme performance when the beamline parameters are changed. Moreover, such an approach
does not allow one to predict the bandwidth of the resulting bunching if the initial modulation is not perfectly coherent. As a result, it is not clear whether the proposed schemes are suitable for FEL
seeding. In this paper we address this drawback and develop general formalism 
quantifying modulation parameters and describing evolution of bunched beams along an arbitrary linear beamline.
This approach will simplify the analysis and will serve as a powerful tool for designing FEL seeding schemes.

The modulated beam has fine scale features in the phase space distribution. Low order moments of the beam distribution such as rms beam sizes, rms energy spread, {\it etc}. cannot capture these small scale
variations. As a result, one needs to introduce high order moments to characterize the distribution function. That significantly complicates the analysis since the number of parameters characterizing the 
phase space distribution
increases by many orders of magnitude. Alternatively, one can consider the 6-dimensional (6D) Fourier transform of the beam distribution function which introduces spectral distribution function. Such a
spectral distribution is well localized in some region of the spectral domain and it can be characterized with a small number of parameters such as the wavevector of modulation and its bandwidth. These
parameters capture the main properties of the modulation similar to the rms beam quantities characterizing its envelope. This approach is very similar to description of laser pulses which are commonly characterized with the carrier frequency and bandwidth (low order moments of the spectral power distribution).
The spectral distribution changes when the electron beam passes through beam optics elements. As it will
be shown bellow, one can easily find how the spectral distribution evolves along the beamline. That allows finding the change of the rms spectral parameters and eventually describing the evolution of the
bunched beam. 

The paper is organized as follows. In Sec.~\ref{sec:mod} we develop general principles for describing dynamics of modulated beams.
First, we introduce spectral distribution function describing beam distribution
in the spectral domain and derive equation describing its change along linear beamlines. Then we introduce low order moments of the spectral distribution and find how they evolve along the beamline. Finally,
we find several invariants relating spectral properties of modulated beams which remain constant under arbitrary linear symplectic transforms. In Sec.~\ref{sec:modulators} we describe the main methods for
modulating beams and find the resulting spectral distribution function which is generated in these modulators. In Sec.~\ref{sec:schemes} we apply developed formalism for analyzing common schemes for XFEL
seeding. We demonstrate that developed theory significantly simplifies analysis and estimate the bandwidth of the output bunching in case of significant random phase noise of utilized laser pulses.

\section{General description of modulated beams}
\label{sec:mod}

Any electron beam can be described as an ensemble  of electrons occupying some phase space volume.
Each electron can be fully described by its position in the 6D phase space
\be
\label{zeta}
{\bf \zzeta}(s)=(x, p_x, y, p_y, \Delta t, -\Delta E),
\ee
where $x$ and $y$ are the transverse coordinates of electron in respect to the reference trajectory, $p_x$ and $p_y$ are the corresponding canonical momenta, $\Delta t$ is the deviation of arrival time
to position $s$ along the beamline, $\Delta E$ is the deviation of the particle energy from the average bunch energy. 

The entire bunch as an ensemble can be described with the distribution function $f(\zzeta)$ which characterizes electron density in the 6D phase space.
We anticipate describing modulated electron bunch which manifests as fine scale variations in the phase space. Describing such a beam with rms quantities will require finding high order
correlations in order to capture short scale variations. Therefore, it is more practical to describe modulated beams with the entire distribution function rather than with a very large number of high order
moments.

\subsection{Vlasov equation}
\label{sec:Vlasov}

Along the beamline electrons interact with electro-magnetic fields which satisfy Maxwell equations.
That implies that the forces can be described with Hamiltonian $H(\zzeta,s)$ which describes the entire dynamics. The
evolution of the distribution function satisfies Vlasov equation which can be considered as a continuity equation in the phase space
\be
\label{Vlasov_general}
\frac{df}{ds}=\partial_sf(\zzeta,s)+[f,H]=0,
\ee
where $[f,H]=(\nabla f)^TJ(\nabla H)$ is the Poison bracket, $J$ is the unit block-diagonal antisymmetric symplectic matrix, $\nabla$ is
the 6D gradient in the phase space, and superscript $T$ stands for transposition.

Vlasov equation is a hyperbolic partial differential equation and it can be solved using characteristics method. Each characteristic describes electron trajectory in the phase space and generalizes
Newton equations for an arbitrary choice of canonical variables $\zzeta$:
\be
\label{Newton}
\frac{d\zzeta}{ds}=J(\nabla H).
\ee
In this paper we neglect that particles interact with each other (Hamiltonian does not depend on the distribution function).
We also limit our analysis to linear beamline optics. Such a system can be described with Hamiltonian $H$ which quadratically depends on the phase space coordinate $\zzeta$.
The Hamiltonian can be represented as a quadratic form defined by the symmetric matrix $\mathcal{H}=\mathcal{H}^T$
\bea
\label{Ham_quad}
&&H(\zzeta,s)=\frac{1}{2}\zzeta^T\mathcal{H}(s)\zzeta,\\
\label{Vlasov_linear}
&&\frac{df}{ds}=\partial_sf(\zzeta,s)+(\nabla f)^TJ\mathcal{H}\zzeta=0,
\eea
Under this assumption, 
Newton equations (\ref{Newton}) become linear
\be
\label{Newton_linear}
\frac{d\zzeta}{ds}=J\mathcal{H}\zzeta
\ee
and they can be solved using, for example, Magnus expansion. The final and initial coordinates of each particle transform linearly by
conventional symplectic transform matrix $R$. Then the formal solution of Vlasov equation 
\bea
\label{zeta_transform}
&&\zzeta(s)=R(s,s_0)\zzeta(s_0),\\
\label{Vlasov_solution}
&&f\left(\zzeta,s\right)=f\left(R^{-1}(s,s_0)\zzeta,s_0\right).
\eea
This solution represents the Liouville theorem which states that the particle density is conserved along the trajectory in the phase space.

\subsection{Spectral distribution function}
\label{sec:Vlasovk}

As mentioned above, the distribution function $f(\zzeta)$ of the modulated beam has small scale variations and it cannot be characterized with a small number of typical parameters. However, these variations
are quasi-periodic. Therefore, the Fourier spectrum of the distribution function is well localized and it is more uniform than the phase space distribution function.
It can be used then for description of modulated beams. This approach is similar to methods
used in Optics where laser pulses are conventionally described with their spectra rather than the actual field distribution in space. Unlike Optics, the beam is characterized with the distribution function
in the 6D phase space and any component of the electron 6D position can change along the beamline. Therefore, one has to consider 6D Fourier transform of the distribution function which introduces 
spectral distribution function
\be
\label{fk}
f_{\bf k}({\bf k},s)=\int f(\zzeta,s)e^{-i{\bf k}^T\zzeta}d^6\zzeta.
\ee

The spectral distribution function fully describes the beam since the phase space distribution
can be recovered through inverse Fourier transform of the spectral distribution. At the same time, the spectral distribution
function reflects important characteristics of the beam which are not evident from the distribution function $f(\zzeta)$ in the phase space. For example, spectral distribution function along
longitudinal axis $\hat{\kk}_5\equiv\hat{\kk}_z$
is equal to the beam bunching factor, $b(k)=\int f(\zzeta)e^{-ik\zeta_5} d^6\zzeta\equiv f_\kk(\kk=\hat{\kk}_5k)$.

Description of modulated beams with the spectral distribution function is beneficial only if its change along the beamline can be described with simple enough equations. Otherwise, the model will
be too complicated for practical use. Bellow we find the exact equation describing the change of the spectral distribution but first we show that its dynamics is simple enough. Consider a single harmonic of
the spectral distribution function $f_\kk(\kk,s_0)=\delta(\kk-\kk^\prime)$ which corresponds to the distribution function $f(\zzeta,s_0)\propto\exp(i\kk^{\prime T}\zzeta)$. The phase of the modulation
$\phi=\kk^T \zzeta$ linearly depends on the phase space coordinate $\zzeta$. Therefore, the phase of the modulation remains linear under any linear transform of the phase space $\zzeta\rightarrow R \zzeta$,
which is the case of linear beamline optics. Therefore, a monochromatic modulation transforms into another monochromatic modulation having different wavevector, $\kk^\prime\rightarrow\kk^{\prime\prime}$.
As a result, different harmonics of the spectral distribution do not mix with each other in a linear beamline and spectral distribution function is self-similar along the beamline. This property indicates
that the equation describing evolution of the spectral distribution function should be simple enough.

We find the precise equation describing dynamics of the spectral distribution function by taking Fourier transform of Vlasov equation (\ref{Vlasov_general}) assuming
quadratic form of the Hamiltonian described by Eq.~(\ref{Ham_quad}). Using equality $\tr(J\mathcal{H})=0$ one can end up with the following equation after some straightforward algebra 
\bea
\nonumber
\frac{df_{\bf k}}{ds}&=&\partial_sf_{\bf k}+[f_\kk,H_\kk]_\kk=\\
\label{Vlasov_k}
&=&\partial_sf_{\bf k}+(\nabla_{\bf k}f_{\bf k})^T\mathcal{H}J{\bf k}=0,\\
\label{Hk}
H_\kk(\kk,s)&=&-\frac{1}{2}\kk^T J\mathcal{H}(s)J\kk.
\eea
This equation has exactly the same form as the original Vlasov equation (\ref{Vlasov_linear}) and can be interpreted then as Vlasov equation for the spectral distribution function.
The dynamics of the spectral distribution function is symplectic and it is fully described with Hamiltonian $H(\kk,s)$ defined by Eq.~(\ref{Hk}).

According to Liouville theorem the spectral distribution function does not change along the trajectory in $\kk$ space, and therefore, it can be considered as an ensemble of non-interacting quasi-particles. Each
quasi-particle represents one spectral harmonic of the spectral distribution. These quasi-particles do not vanish or born in time, they do not interact or mix with each other, and their entire dynamics 
manifests as motion along trajectories in $\kk$ space. The only fundamental difference between descriptions in the phase space and spectral domains
is that the spectral distribution function can be complex unlike the phase space distribution function which is real and positive. 

Each harmonic of the spectral distribution function remains as a single harmonic along linear beamline. This property indicates that linear beam optics cannot be used for generating
high order harmonics of modulation.
This observation does not contradict performance of the HGHG \cite{HGHG} or EEHG \cite{EEHG} schemes.
As it will be shown in Sec.~\ref{sec:laser} high order harmonics are generated in the modulator in which the phase space of the beam transforms nonlinearly.
The following chicanes do not create additional harmonics in the spectral domain but only transform imposed modulation into longitudinal bunching.

In order to solve spectral Vlasov equation (\ref{Vlasov_k}) we note that it is hyperbolic partial differential equation, same as conventional Vlasov equation (\ref{Vlasov_linear}).
Therefore, it can be solved using characteristics method. Each spectral component remains constant along the following trajectory in $\kk$ space
\be
\label{Newton_k}
\frac{d{\bf k}}{ds}=\mathcal{H}J{\bf k}.
\ee

Characteristics equation (\ref{Newton_k}) for each spectral harmonic is similar to characteristics equation (\ref{Newton_linear}) describing trajectory of each individual particle. Therefore, these linear
equations can be solved using the same techniques. At the same time, these two solutions are related to each other since corresponding equations contain the same matrices. One can note that two arbitrary
solutions of Eqns.~(\ref{Newton_linear}) and (\ref{Newton_k}) do not evolve independently and their product is invariant along the beamline
\be
\label{Newton_kz}
\frac{d({\bf k}^T\zzeta)}{ds}=0,
\ee
Using solution for particle trajectory (\ref{zeta_transform}) and noticing that relation (\ref{Newton_kz}) holds for any arbitrary
initial particle position $\zzeta(s_0)$, one finds the solution for the spectral characteristics
\be
{\bf{k}}(s)=R^{-T}(s,s_0){\bf k}(s_0).
\ee
This relation describes the change of each spectral component wavevector. The solution of the spectral Vlasov eqaution (\ref{Vlasov_k}) reads then as
\be
\label{solution_fk}
f_{\bf k}({\bf k},s)=f_{\bf k}(R^{T}(s,s_0){\bf k},s_0).
\ee

\subsection{Spectral averages}
\label{sec:stat}

The beam is fully described with its spectral distribution function $f_\kk(\kk,s)$. At the same time, few average parameters of the spectral distribution function 
may suffice to characterize its main features. We introduce the spectral averaging of some variable $g({\bf k})$ as
\be
\label{averages}
\overline{ g(\kk)}(s)=\frac{\int g(\kk)|f_\kk(\kk,s)|^2 d^6\kk}{\int |f_\kk(\kk,s)|^2d^6\kk}.
\ee
Note that we use overline notation $\bar{\cdot}$ to characterize spectral averages in order to distinguish them from the averages in the phase space domain defined as $<\cdot>$. The
denominator in Eq.(\ref{averages}) remains constant in linear beamlines and serves as the normalization factor so that the quantity itself and its spectral average have the same dimensions.

In case of narrowband modulations the spectral distribution function is localized in some region of the spectral domain. Such a distribution can be well characterized with few low order moments.
We introduce the spectral average wavevector of modulation $\overline{\kk}(s)$ and the second order spectral correlation matrix
\be
B(s)=\overline{(\kk-\overline{\kk})(\kk-\overline{\kk})^T}.
\ee
This matrix describes rms spreads of the spectrum distribution function $f_\kk(\kk,s)$. The diagonal elements describe the modulation bandwidths along the corresponding axes. Therefore, we call matrix $B$ as
the ``bandwidth matrix". The bandwidth matrix defines the rms ellipsoid in the spectral domain similar to the rms envelope ellipsoid defined by the beam matrix $\Sigma=<\zzeta\zzeta^T>$.
This rms bandwidth ellipsoid qualitatively shows noise distribution along different dimensions.

In some cases the spectral distribution function $f_\kk(\kk,s)$ can be localized in several well separated regions of the spectral domain,
{\it e.g.} the spectrum of the beam modulated with the laser pulse consists of several localized harmonics as will be shown in Sec.~\ref{sec:laser}. 
Vlasov equation (\ref{Vlasov_k}) for the spectral distribution function states that different trajectories in the spectral domain do not cross which does not allow well 
separated spectral domains
mixing with each other. In this case the spectral average quantities can be introduced for each localized domain individually.

Using solution (\ref{solution_fk}) for the spectral distribution function along the beamline, one can easily find transforms of the spectral averages
\bea
\label{k_transform}
&&\overline{\kk}(s)=R^{-T}(s,s_0)\overline{\kk}(s_0),\\
\label{B}
&&B(s)=R^{-T}(s,s_0)B(s_0)R^{-1}(s,s_0).
\eea

Note that the wavenumber of modulation transforms linearly along the beamline. This property indicates that scaling the wavenumber of modulation without changing the wavevector orientation results in the
same scaling
of the output modulation wavenumber. This property can be used, for example, in the masked dogleg \cite{Brookhaven_dogleg} or Emittance EXchanger (EEX) \cite{EEX_exp} 
schemes to generate bunches with different spatial spacing. This task can be accomplished by using several masks with different spacing between the slits. At the same time, the optics required
to transform imposed modulation into longitudinal bunching remains the same for different masks.

The bandwidth matrix $B(s)$ transforms exactly in the same way as the inverse beam matrix $\Sigma^{-1}$,
\be
\label{Sigma_transform}
\Sigma(s)\equiv <\zzeta \zzeta^T>
=R(s,s_0)\Sigma(s_0)R^T(s,s_0).
\ee
This property becomes evident if one considers unmodulated
beam with some envelope. In this case the average modulation wavenumber is close to zero and the bandwidth matrix describes the rms wavevector spread of the spectral distribution which is inverse proportional
to the rms beam sizes. Moreover, if one considers the beam with 6D Gaussian distribution, $f\propto \exp(-\zzeta^T\Sigma^{-1}\zzeta/2)$, then the bandwidth matrix is related to the rms beam matrix as
$B=\Sigma^{-1}/2$ (this property can be proved through direct calculation of integrals in Eqs.~(\ref{fk}) and (\ref{averages}) by transforming integration variables according to Eq.~(\ref{diag})
which diagonalizes matrices). This relation also holds if arbitrary monochromatic modulation is imposed on the beam with 6D Gaussian envelope since this modulation changes the average wavevector but does not
affect the bandwidth matrix. At the same time, if the imposed modulation is not monochromatic and has some bandwidth due to noise, then the resulting bandwidth matrix deviates from $\Sigma^{-1}/2$.
This example shows that describing modulated beam with the bandwidth matrix is useful only in the presence of significant noise when $||I-2B\Sigma||\gg ||I||$. Otherwise, the modulation
can be considered monochromatic and it is more convenient to describe the beam with its phase space envelope and a single wavevector of modulation. Introducing the bandwidth matrix for the monochromatic
modulation is not feasible since it is strongly related to the beam matrix $\Sigma$ and does not provide additional information.

Note that the bandwidth matrix is a useful concept for describing the beam quality even if the beam is not modulated. While the rms beam matrix $\Sigma$ describes large scale features of the beam such as its
rms sizes, energy, and angular spreads, the bandwidth matrix $B$ describes small-scale fluctuations in the distribution function. Therefore, the bandwidth matrix describes the beam homogeneity and
it is more homogeneous at smaller values of $||I-2B\Sigma||$.

The modulation wavevector and the bunch envelope transform independently from each other since their transforms depend only on the transform matrix $R$ but not on each other.
Identical envelopes transform in the same wave regardless what kind of modulation is imposed on the bunch. Identical modulations imposed on bunches
with different envelopes also transform in the same way. This property significantly simplifies
the task of designing beam buncher since it can be designed independently from the rest of the beamline. This beamline section should solve the problem of transforming imposed modulations into required
modulations and this optics does not depend on rms beam quantities such as its sizes and emittances. Once designed, this block can be attached to any accelerator and it will perform
equally in any regime of the machine.

\subsection{Modulation invariants}
\label{sec:inv}

The bandwidth matrix is a positively defined symmetric matrix which transforms along the beamline through symplectic matrix $R^TJR=J$ since it represents Hamiltonian dynamics.
This transform is similar to the transform (\ref{Sigma_transform})
of the beam rms matrix $\Sigma$. Therefore, the invariants which hold for the beam matrix \cite{Dragt_eig} are also applicable for the bandwidth matrix
\bea
&& \det(B)=inv,\\
&& \tr(BJ)^{2n}=inv,\;\;\;\;\;n=1,2,\,...\,dim(B)/2.
\eea
These invariants can be used to introduce parameters similar to eigen-emittances for the beam envelope \cite{Dragt_eig}.

Other than that, one can note that the change of the spectral averages is fully described with the transform matrix $R$. This is not surprising since the transform matrix defines the mapping of each particle
in the phase space, and therefore, describes the entire beam dynamics. Therefore, spectral averages do not transform fully independently from the beam rms quantities. Combining Eqs.~(\ref{k_transform}) and
(\ref{Sigma_transform}) one can easily find the following scalar quantity which remains constant along arbitrary linear beamline
\be
\label{inv_Phi}
\overline{\kk}^T(s)\Sigma(s)\overline{\kk}(s)=inv.
\ee 
This quantity can be interpreted as the overall phase of the modulation across the bunch envelope. To illustrate this invariant one may consider imposing longitudinal bunching on the beam and its
further compression. In this case, the number of modulation periods within the rms beam length remains constant during compression, {\it i.e.} the overall phase of modulation is invariant.

Additional invariants can be found by taking into account that transform matrix $R$ describes Hamiltonian dynamics and, therefore, is symplectic.
This condition limits possible changes of the beam matrix and modulation wavenumber. Using Eqs.~(\ref{k_transform}), (\ref{Sigma_transform}), and taking into account that transform matrix $R$ is symplectic
one can construct the following invariants
\bea
\label{inv}
&\overline{\kk}^TJ(\Sigma^{-1}J)^{2n+1}\overline{\kk}\equiv \overline{\kk}^T(J\Sigma^{-1})^{2n+1}J\overline{\kk}=inv,&\\
\nonumber
&n=1,2,\,...\,dim(\Sigma)/2.&
\eea
There is infinite number of invariants but the number of functionally independent ones is equal to the dimension of space describing beam dynamics, {\it i.e.} it is equal to 3 in general case. Moreover,
these invariants are not functionally independent from Eq.~(\ref{inv_Phi}).

Invariants (\ref{inv}) can be understood using eigen-emittance concept as described in Appendix~\ref{app:inv}. They state that the overall
phase of modulation across each eigen- phase plane is preserved. This property can guide attempts for beam bunching using transverse masking.
It is of particular practical interest to consider beams which eigen-emittances are recovered as $x$-, $y$-, and $z$- emittances
since this scenario corresponds to the brightest beam with the fixed phase space volume.
Transverse masking modulates beam only in a single eigen- phase plane (which coincides with $x$- phase plane) in case of
uncorrelated beam. According to invariants (\ref{inv}) the modulation imposed in one eigen- phase plane remains in the same eigen- phase plane along the beamline. This indicates that transform of
$x$-modulation into $z$-modulation should be accompanied with transform of $x$- eigen- phase plane into $z$- eigen- phase plane. Therefore, one should use EEX optics \cite{EEX} to achieve this goal.
Otherwise, the modulation can be smeared out at significantly large emittances as observed in Ref.~\cite{Brookhaven_dogleg}.

The bandwidth matrix transforms exactly in the same way as the inverse rms beam matrix. Therefore, the invariants involving the bandwidth matrix are similar to invariants (\ref{inv_Phi}) and (\ref{inv})
found for the beam matrix
\bea
\label{bandwidth}
&&\overline{\kk}^TB^{-1}\overline{\kk}=inv,\\
&&\overline{\kk}^TJ(BJ)^{2n+1}\overline{\kk}=inv,\;\;\;\;
n=1,2,3.
\eea
Invariant described with Eq.~(\ref{bandwidth}) can be interpreted as the conservation of relative bandwidth $\delta k/k$ along the beamline. In case of homogeneous modulations, invariant (\ref{bandwidth})
is exactly the same as invariant (\ref{inv_Phi}) since the bandwidth matrix is proportional to the inverse beam matrix and the relative modulation bandwidth is roughly equal to the inverse number of
modulation periods per bunch.

\section{Beam modulators}
\label{sec:modulators}
In this section we describe ways to impose modulations on the beam. Currently, there are two known ways for modulating beams, namely transverse masking and imposing spatially dependant energy
modulation.

\subsection{Transverse masking}
\label{sec:masking}
The beam can be modified by passing it through the mask. The mask serves as a filter which either absorbs some particles or strongly changes their phase space coordinates so that they can be filtered
downstream the beamline. Significant change of the phase space coordinates for some group of particles while keeping intact the rest of the beam requires presence of non-Hamiltonian forces such as
incoherent damping, scattering, ionization, {\it etc}. The simplest mask can be envisioned as a perforated foil installed transversally to the bunch propagation. Electrons passing through the foil strongly
scatter and are lost downstream while electrons passing through the opening remain intact. These masks modify the bunch transversally and they are widely used in accelerator technology.
Modifying the distribution function along other phase space coordinates would require differentiation of particles based on their position in phase space:
different ionization cross-section at different
energies, triggering interaction externally at a certain time, {\it etc}.

Without addressing particular physical mechanism of masking, we consider that the bunch distribution function can be changed by the mask as
\be
f(\zzeta)={f^0}(\zzeta)M(\zzeta),
\ee
where ${f^0}$ is the distribution function of the bunch before the mask and $M$ is the the mask imprint.
The spectral distribution function of the resulting beam $f_{\bf k}(\kk)$ can be presented as a convolution of the initial and the mask spectral distribution functions $f^0_{\bf k}(\kk)$ and $M_{\bf k}(\kk)$,
respectively
\be
\label{mask_spec}
f_{\bf k}({\bf k})=f_\kk^0*M_\kk=\int f^0_{\bf k}({\bf k^\prime})M_{\bf k}({\bf k-k^\prime})d^6{\bf k^\prime},
\ee

This equation provides a semi-qualitative description for the modulated beam spectral properties. We consider both the beam and the mask spectra to be well-localized.
If their bandwidths can be neglected {\it i.e.} their spectra can be well
approximated with $\delta$-functions, the wavevector of the resulting modulation can be approximated as
\be
\overline{\kk}[f_\kk]\approx \overline{\kk}[f^0_\kk]+\overline{\kk}[M_\kk].
\ee
Note that using perforated foil as a mask modulates the beam in the transverse plane. For example,
a set of wires aligned along $y$ and placed periodically along $x$ would shift the average wavevector of modulation along $k_x$.

Taking into account finite bandwidths of the initial beam and the mask results in the finite bandwidth of the resulted modulation.
In some cases (Gaussian or uncorrelated spectra) the spectral bandwidths add in quadrature to each other. In general case, the bandwidth of the resulting beam cannot be related to the bandwidths of the
initial beam and the mask without knowledge about the spectral distribution functions. It can be just noted that if one of the bandwidth is much larger then another, then the bandwidth of the
convolution is approximately equal to the largest one.

\subsection{Energy modulation with laser pulse}
\label{sec:laser}
In this scheme the beam is modulated by applying spatially dependent energy modulation \cite{HGHG}. Such a modulation can be imposed by passing a beam through the undulator and its interaction with resonant
radiation. Interaction of the laser pulse with electron beam results in the change of the particle energy $E\rightarrow E-\delta E\sin\varphi(z)$. The resulting beam distribution function $f(z,E)$ relates
to the distribution function of unperturbed beam $f^0(z,E)$ as
\be
\label{f_laser}
f\left(z,E\right)={f^0}\left(z,E+\delta E \sin\varphi(z)\right),
\ee
where for simplicity of description we omitted transverse variables.

Ideally, the laser is considered to be monochromatic, {\it i.e.} $\varphi(z)=kz$. In order to extend our analysis we consider that the laser has finite bandwidth caused by random fluctuations in its phase
(and consider the laser amplitude to be spatially uniform). As a result, the laser phase has linearly growing and random components. In order to characterize such a laser, we consider average quantities
of its spectrum
\bea
&&\varphi_k(k)=\int e^{i\varphi(z)-ikz}dz,\\
&&\overline{k_\varphi}=\overline{k}[\varphi_k],\;\;\;\;\;\;\;\overline{\delta k_\varphi^2}=\overline{(k-\overline{k_\varphi})^2}[\varphi_k].
\eea

\begin{figure}[ht]
\includegraphics[width=3.4in,keepaspectratio]{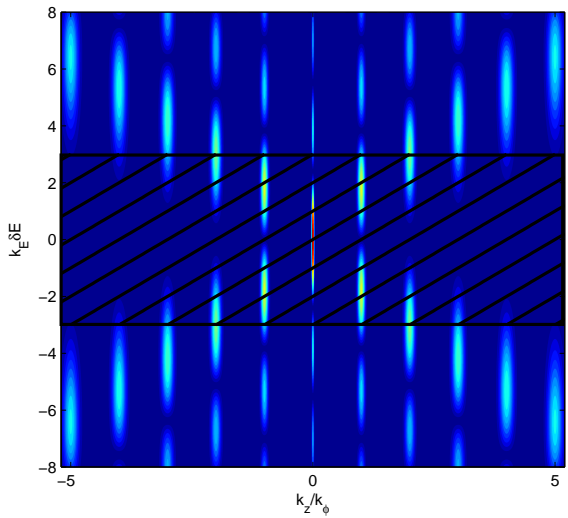}
\caption{(Color online) Spectrum of modulation produced by spatial variation of beam energy. Dashed area schematically shows the energy bandwidth of the beam envelope.}
\label{energy_mod}
\end{figure}

The spectral distribution function of the modulated beam described with Eq.~(\ref{f_laser}) can be related to the spectral distribution functions of the initial beam $f_\kk^0$
and the spectral distribution of the laser phase $\varphi_k$. After some straightforward algebra one can find that the beam spectrum consists of infinite number of harmonics
\bea
\label{mod_laser}
&&f_\kk(k_z,k_E)=\sum\limits_{n=-\infty}\limits^\infty f_\kk^{(n)}(k_z,k_E)=\\
\nonumber
&&=\sum\limits_{n=-\infty}\limits^\infty J_n(|k_E\delta E|)
f_\kk^0(k_z,k_E)*\underbrace{\varphi_k(k_z)*...*\varphi_k(k_z)}_n,
\eea
where $J_n$ is the $n$-th order Bessel function. The schematics of the spectrum is presented in Fig.~\ref{energy_mod}.

The form of the spectral distribution function described with Eq.~(\ref{mod_laser}) determines the main parameters of the modulation spectrum. The spectrum consists of the infinite number of harmonics. The 
spectral distribution function along longitudinal wavenumber $k_z$ is described with $n$ times convolution of the laser spectral distribution $\varphi_k(k_z)$. As discussed in Sec.~(\ref{sec:masking}) the
spectral properties of such a distribution cannot be found in general case. However, if one assumes that the laser phase noise is stationary and Gaussian (the laser spectrum is $\delta$-correlated
which can be assumed if the laser bandwidth is mainly defined by the noise rather than its envelope and the spectrum statistics is Gaussian) then the spectral properties of the beam modulation can be
found as
\bea
&&\overline{k_z}[f_\kk^{(n)}]= \overline{k_z}[f^0_\kk]+n \overline{k_\varphi},\\
&&B_{k_z,k_z}[f_\kk^{(n)}]= B_{k_z,k_z}[f^0_\kk]+|n|\overline{\delta k_\varphi^2},
\eea
{\it i.e.} the carrier frequency scales linearly with the harmonic number and the bandwidths add in quadrature.
At the same time, if the laser phase is not random then the modulation bandwidth increases much faster with harmonic number. For example, if one considers chirped laser pulse then
the modulation bandwidth would scale linearly with harmonic number, $B_{k_z,k_z}=B_{k_z,k_z}[f^0_\kk]+n^2\overline{\delta k_\varphi^2}$. This effect
can be proved by expressing $n-$th order convolution as $\varphi_k(k_z)*...*\varphi_k(k_z)=\int e^{i n\varphi(z)-ik_zz}dz$, which indicates linear scaling of the chirp with harmonic number.

Unlike the longitudinal spectral distribution along $k_z$, the energy spectral distribution along $k_E$ is equal to the product of the initial beam spectrum and the
modulation spectrum which is equal to the Bessel
functions. The product of these two spectra does not vanish as long as both distributions are significant in the same domain. The schematics of the beam spectrum resulted from its energy modulation with laser
is shown in Fig.~\ref{energy_mod}. Dashed area schematically shows the energy bandwidth caused by the initial beam envelope. The harmonics having significant amplitude within that bandwidth are
present in the spectrum and high order harmonics are heavily suppressed.

In order to find the energy wavenumber $({k}_E^{(n)})_{\rm max}$ which maximizes each harmonic one needs to maximize the spectral distribution function
\be
\label{harmonic_max}
\left.\frac{\partial}{\partial_{k_E}} \left(J_n(|k_E\delta E|)f_\kk^0(k_z,k_E)\right)\right|_{(k_E^{(n)})_{\rm max}}=0.
\ee
This equation can be solved assuming certain beam energy distribution, otherwise the problem
cannot be solved analytically.
Typically the beam is assumed to have Gaussian energy distribution (so as its energy spectrum) \cite{HGHG,EEHG,CHG_Ratner}.
The main features can be recovered considering beam energy bandwidth to be very large, {\it i.e.} the beam energy spread is considered to be small. Then
the maximum of $n$-th harmonic is not affected by the energy bandwidth of the original beam envelope and it is located at $({k}_E^{(n)})_{\rm max}\approx\mu_{n1}/\delta E,$
where $\mu_{n1}$ is the first maximum of $n$-th order Bessel function, $J^\prime_n(\mu_{n1})=0$. Therefore, the modulation is not heavily suppressed by the finite beam energy spread if the modulation is
significantly large $({k}_E^{(n)})_{\rm max}\lesssim (B_{k_E,k_E}[f_\kk^0])^{1/2}$ or $\delta E\gtrsim\mu_{n1}\sigma_E\sim n \sigma_E$, where $\sigma_E$ is the rms energy spread of the original beam envelope.
Note that this condition is harder to satisfy for modulations at higher harmonics which agrees with scalings of HGHG scheme \cite{HGHG}.

The energy bandwidth of the resulting distribution ($B_{k_E,k_E}$ element of the bandwidth matrix) is not well defined since the spectral energy distribution consists of several spikes which are described
with oscillating Bessel function. One cannot consider them individually since they are not well separated in the spectral domain.
However, one can characterize the spectral distribution bandwidth with its curvature at
the maximum, $B_{k_E,k_E}\sim-0.5f_\kk(k_z,k_E)/\partial^2_{k_E,k_E}f_\kk(k_z,k_E)$. For low order harmonics, $\mu_{n1}\lesssim\delta E /\sigma_E$, one finds
\bea
\nonumber
&&B_{k_E,k_E}^{(n)}\approx-\frac{1}{2}\left.\frac{f_\kk^{(n)}(k_z,k_E)}{\partial^2_{k_E,k_E}f_\kk^{(n)}(k_z,k_E)}\right|_{(k_E^{(n)})_{\rm max}}=\\
\label{Bkeke}
&&=\frac{\mu^2_{n1}}{\mu^2_{n1}-n^2}\frac{1}{(\delta E)^2}\sim\frac{0.618n^{2/3}}{(\delta E)^2},\;\;\;\; 
\begin{array}{l}
n\gg1 \\
n\lesssim\delta E /\sigma_E
\end{array}
\eea

Eq.~(\ref{mod_laser}) shows that the non-diagonal terms of the resulting bandwidth matrix $B_{k_z,k_E}=B_{k_E,k_z}$ are equal to zero if these terms were zero for the original beam envelope before the
modulation was applied. This approximation is typically valid for beam without energy chirp. Then the bandwidth matrix of the modulated beam can be approximated as
\be
\label{band_laser}
B[f_\kk^{(n)}]=\left(
\begin{array}{cc}
B_{k_z,k_z}[f^0_\kk]+|n|\overline{\delta k_\varphi^2} & 0\\
0& \frac{\mu^2_{n1}}{\mu^2_{n1}-n^2}\frac{1}{(\delta E)^2}
\end{array}\right).
\ee

Note that this expression describes characteristic spectral bandwidth rather than exact spectral average value. Therefore, transforms and invariants for the bandwidth matrix found in Sec.~\ref{sec:stat}
and \ref{sec:inv} should be used with caution.

\section{Transforms of beam modulation in various schemes}
\label{sec:schemes}

In this Section we illustrate the developed formalism describing evolution of modulated beams in the spectral domain. We consider common schemes for XFEL seeding which rely on beam modulation by laser
and transforming imposed modulation into longitudinal bunching at small wavelengths.
As described in Sec.~\ref{sec:laser} the beam dynamics occurs only in the longitudinal phase space, $\zzeta=(t,-\Delta E)^T$.
Therefore, we limit our analysis to the 2D distribution function in the longitudinal phase space and its spectrum.

Conventional schemes for creating beam bunching from the laser-induced energy modulation rely on two types of beam optics elements, namely chicanes and RF cavities introducing energy chirp. These elements
transform the wavevector of modulation in 2D spectral domain as $\kk=R^{-T}\kk^0$, where the transform matrices for these elements have the following form:
\be
R_{\rm chicane}=\left(
\begin{array}{cc}
1 &\xi\\
0& 1\\
\end{array}\right)\;\;\;\;\;\;\;\;\;\;\;
R_{\rm cavity}=\left(
\begin{array}{cc}
1 &0\\
\alpha& 1\\
\end{array}\right).
\ee
As a result, the wavevector of modulation transforms by these elements as
\bea
\label{chicane}
\left(
\begin{array}{c}
k_z\\k_E
\end{array}
\right)=&R^{-T}_{\rm chicane}
\left(
\begin{array}{c}
k_z^0\\k_E^0
\end{array}
\right)
&=\left(
\begin{array}{c}
k_z^0\\k_E^0-\xi k_z^0
\end{array}
\right),\\
\label{cavity}
\left(
\begin{array}{c}
k_z\\k_E
\end{array}
\right)=&R^{-T}_{\rm cavity}
\left(
\begin{array}{c}
k_z^0\\k_E^0
\end{array}
\right)
&=\left(
\begin{array}{c}
k_z^0-\alpha k_E^0\\k_E^0
\end{array}
\right).
\eea

It is easy to follow these transforms in the 2D spectral plane similar to what was presented in Fig.~\ref{energy_mod}. Any given scheme can be represented with a diagram which shows
trajectory of the modulation wavenumber between the initial and final values. The following notations on these diagrams will be used:
\begin{itemize}
\item Consider the laser induced modulation imposed on the beam.
This modulation has spectral distribution function which peaks at the wavevector $\kk=(2\pi n/\lambda,(k_E^{(n)})_{\rm max})^T$, where $\lambda$ is the laser wavelength and
$(k_E^{(n)})_{\rm max}$ can be found from Eq.~(\ref{harmonic_max}). We will focus on the evolution of the vicinity of this spectral component
since it has the largest harmonic current and we will mark this wavevector with a cross. 

\item Each beam optics element
transforms the modulation wavevector according to Eqs.~(\ref{chicane}) and (\ref{cavity}). The modulation wavevector changes its $k_E$ component when the beam passes through some chicane and it
changes $k_z$ component when the beam passes through some RF cavity. We will illustrate these transforms with arrows starting from the initial and ending at the final wavevectors of modulations. Note
that chicanes and cavities will be represented as  arrows parallel to the spectral space axes. Also note that chicanes are typically considered to have positive dispersion $\xi>0$. Therefore, they will be represented
with downward arrows in the right-hand side of the diagram (positive harmonics numbers, $k_z>0$) and upward arrows in the left-had side of the diagram (negative harmonics numbers, $k_z<0$).

\item Some schemes require imposing laser-induced modulation several times along the beamline. As discussed in Sec.~\ref{sec:laser} the modulation does not vanish only in the vicinity of $k_E$ where both the
beam envelope and the modulation have significant spectral distributions. In these schemes the harmonic current imposed by the first laser and transformed in the beamline can serve as an envelope in
the following modulator. We will mark this mechanism by a dashed arrow parallel to $k_z$ axis (modulation by the second laser does not vanish only in the vicinity of $k_E$ of already existing
modulations). Not that this arrow refers to the nonlinear modification of the beam spectrum unlike solid arrows which refer to the linear transforms in which different spectral harmonics do not mix with each
other.
\end{itemize}

All the schemes considered bellow are designed to transform initial modulation into longitudinal bunching. Therefore, the linear beamline optics is aimed to transform the wavevector of the imposed
modulation into $\kk=(k_z,0)^T$, {\it i.e.} to eliminate any energy modulation.

\subsection{High Gain Harmonic Generation (HGHG)}
\label{sec:HGHG}

\begin{figure}[ht]
\includegraphics[width=2.2in,keepaspectratio]{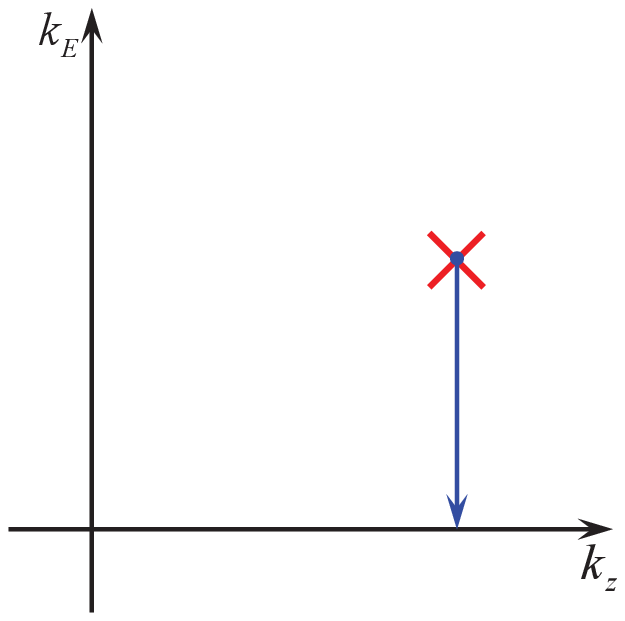}
\caption{(Color online) Schematics for the modulation transform in the HGHG scheme.}
\label{fig:HGHG}
\end{figure}

In this scheme the energy modulation imposed by the laser pulse in a short undulator section is linearly transformed by the following chicane \cite{HGHG}. The schematics for the transform of the modulation
wavevector is presented in Fig.~{\ref{fig:HGHG}}. Considering the final modulation to be longitudinal bunching ({\it i.e.} $k_E=0$), one can easily recover the chicane strength which maximizes the output
bunching factor
\be
\xi=\frac{(k_E^{(n)})_{\rm max}}{2\pi n/\lambda}\sim\frac{\lambda}{2\pi\delta E},\;\;\;\; {\rm for}\;\;\;\;
\frac{\delta E}{\sigma_E}\gg n\gg1\\
\ee

As discussed in Sec.~\ref{sec:laser} and illustrated in Fig.~\ref{energy_mod}, the modulation at high harmonics is heavily suppressed because the $(k_E^{(n)})_{\rm max}\sim n/\delta E$ 
component of the imposed modulation wavenumber increases with harmonic number and eventually does not fit within the energy bandwidth of the beam envelope. That results in reduced amplitude of the spectral distribution
as follows from Eq.~(\ref{mod_laser}). The largest harmonic number which is present in the beam spectrum can be estimated as $n_{\rm max}\sim \delta E/\sigma_E$.

The bandwidth matrix can be found using Eq.~(\ref{band_laser}). Assuming that the bandwidth is mainly determined by the laser phase noise, one obtains the
following bandwidth matrix of the modulation at the end of the HGHG scheme
\bea
\nonumber
B_{\rm HGHG}^{(n)}&=&R_{\rm chicane}^{-T}B[f_\kk^{(n)}]R_{\rm chicane}^{-1}=\\
&=&\left(\begin{array}{cc}
n\overline{\delta k_\varphi^2} & -\xi n \overline{\delta k_\varphi^2}\\
-\xi n \overline{\delta k_\varphi^2}& \frac{\mu^2_{n1}}{\mu^2_{n1}-n^2}\frac{1}{(\delta E)^2}+\xi^2 n \overline{\delta k_\varphi^2}
\end{array}\right)
\eea
The bandwidth matrix transformed by the chicane is not diagonal anymore. This indicates that the rms bandwidth ellipse is not aligned along $k_z$ and $k_E$ axes. Therefore, the extent of this rms bandwidth
ellipse along $k_z$ axis (which defines the bunching bandwidth) can be significantly smaller than the $B_{k_z,k_z}$ bandwidth matrix element. The bandwidth of the resulting bunching can be
found in the limit of large harmonic number, $n\gg1, \mu_{n1}\approx n+0.81n^{1/3}$
\bea
\nonumber
\left(\frac{\overline{\delta k^2}}{\overline{k}^2}\right)_{\rm HGHG}^{(n)}&=&\frac{1}{(n\overline{k_\varphi},0)\left(B_{\rm HGHG}^{(n)}\right)^{-1}(n\overline{k_\varphi},0)^T}=\\
\label{band_hghg}
&=&\frac{1}{n}\frac{\overline{\delta k_\varphi^2}}{\overline{k_\varphi}^2}\left(1+3.23n^{4/3}\frac{\overline{\delta k_\varphi^2}}{\overline{k_\varphi}^2}\right)^{-1}.
\eea
This estimate indicates that the relative bunching bandwidth at $n-$th harmonic is roughly $\sqrt{n}$ times smaller than the relative bandwidth of the laser with $\delta$-correlated phase noise.
The second term in Eq.~(\ref{band_hghg}) shows additional noise filtering due to the presence of the chicane. However, this term is close to unity for low order harmonics ($n\sim5$)
and relatively narrowband lasers ($\sqrt{\overline{\delta k_\varphi^2}/\overline{k_\varphi}^2}\sim10^{-4}$).

\subsection{Echo Enabled Harmonic Generation (EEHG)}
\label{sec:EEHG}

\begin{figure}[ht]
\includegraphics[width=2.5in,keepaspectratio]{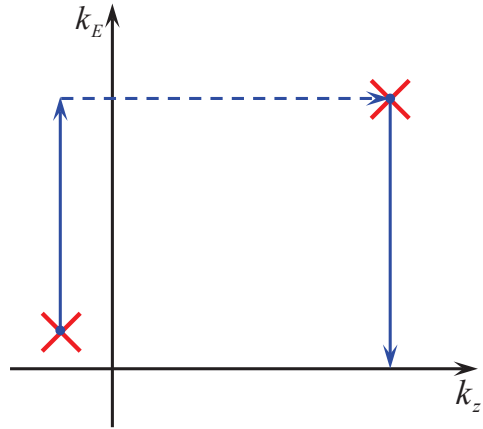}
\caption{(Color online) Schematics for the modulation transform in the EEHG scheme.}
\label{fig:EEHG}
\end{figure}

As discussed in Secs.~\ref{sec:laser} and \ref{sec:HGHG}, the main limitation on generating modulation at high harmonics of the laser wavelength comes from the limited overlap between the energy bandwidths
of the beam envelope and the laser induced modulation. As illustrated in Fig.~\ref{energy_mod}, the modulation at high harmonics is located at high values of $(k_E^{(n)})_{\rm max}\sim n/\delta E$
which eventually fall outside the energy bandwidth of the beam envelope. As a result, the amplitude of the modulation rapidly drops as indicated by Eq.~(\ref{mod_laser}). This drawback can be compensated
by increasing the amplitude of the imposed energy modulation. Such an approach increases modulation amplitude but also results in the increased rms beam energy spread which may reduce
FEL performance. Alternatively, the beam envelope can be modified in such a way that its energy bandwidth is shifted to the domain in which high order harmonics are produced. This is precisely the
scenario which takes place in the EEHG scheme \cite{EEHG}.
In this scheme the beam is first modulated with the laser pulse. The following strong chicane is used to increase the $k_E$ component of the $n_1=-1$ harmonic of
modulation. Then the second modulator is used and $n_1=-1$ harmonic of the already imposed modulation serves as the beam envelope for the secondary modulation. As a result, harmonics are generated in the
domain of increased $k_E$ values, {\it i.e.} high order harmonics with $n_2\gg1$ can be generated. The schematics of the EEHG scheme is presented in Fig.~\ref{fig:EEHG}.

Condition for the overlap of two modulations along $k_E$ coordinate determines the first chicane strength (for the optimized performance of the EEHG scheme at a given harmonic number). The strength
of the second chicane can be determined from the condition that the final modulation should be recovered as the longitudinal bunching, {\it i.e.} $k_E=0$ after the second chicane. Taking these conditions
simultaneously results exactly in the same equations as presented in Ref.~\cite{EEHG}. Note that the first chicane is used to increase $k_E$ component of the beam modulation. This process results in strong
energy modulation of the beam, {\it i.e.} creating energy bands within the beam envelope which are observed in simulations \cite{EEHG}.

The resulting bunching is linearly transformed from the modulation produced in the second modulator. The energy bandwidth $B_{k_E,k_E}$
is mostly determined by the energy bandwidth of $n_1=-1$ harmonic generated in the first modulator for $n=n_1+n_2\gg1$, $\delta E_1\sim \delta E_2$, which follows from Eq.~(\ref{Bkeke}).
At the same time, the spatial bandwidth of the resulting modulation $B_{k_z,k_z}$ increases every time the laser modulation is applied. Finding the
precise bandwidth matrix of the resulting modulation is complicated since the bandwidth matrix is not diagonal before the second modulator due to presence of the first chicane. However,
in the typical regime of small laser bandwidth, $\overline{\delta k_\varphi^2}/\overline{k_\varphi}^2\rightarrow0$, the bandwidth matrix remains mostly diagonal throughout the beamline and one can estimate
the bunching bandwidth generated in EEHG scheme as
\be
\left(\frac{\overline{\delta k^2}}{\overline{k}^2}\right)_{\rm EEHG}^{(n)} \approx \frac{n+2}{n^2}\frac{\overline{\delta k_\varphi^2}}{\overline{k_\varphi}^2},\;\;\;
n\left(\frac{\delta E_1}{\delta E_2}\right)^2\frac{\overline{\delta k_\varphi^2}}{\overline{k_\varphi}^2}\ll1.
\ee
This scaling indicates that the relative bunching bandwidth reduces as $1/\sqrt{n}$ at large harmonic numbers, same as for HGHG scheme.

\subsection{Compressed Harmonic Generation (CHG)}
\label{sec:CHG}
CHG scheme was proposed to generate short wavelength bunching through longitudinal bunch compression \cite{CHG}.
This effect can be understood by considering the invariant (\ref{inv_Phi}). If the beamline is
designed to compress the bunch, the bunching wavelength decreases to keep the number of modulation periods within the bunch fixed. Such a scenario can be realized by passing the beam
through the RF cavity which introduces energy chirp and the following bunch compression in the chicane. As it was noticed in Ref.~\cite{CHG_Ratner},
such a design requires strong RF cavities to introduce significant compression
without smearing out the resulting bunching factor. Alternatively, the resulting bunching smearing can be avoided by decreasing the energy modulation amplitude but then the resulting bunching factor
rapidly drops down same as in HGHG scheme. This effect can be understood by
considering modulation transform in the spectral domain. RF cavity changes the longitudinal component of the beam modulation wavevector $k_z=k_z^0-\alpha k_E^0$
according to Eq.~(\ref{cavity}). The change of the longitudinal
wavenumber is proportional to $k_E$ component of the modulation. That implies that the largest change of the modulation longitudinal wavenumber can be reached at small laser-induced energy modulations
since $k_E\propto 1/\delta E$. At the same time, the amplitude of the laser-induced energy modulation cannot be much smaller than the rms beam energy spread to provide large amplitude of modulations as
follows from Eq.~(\ref{mod_laser}). Therefore, the largest change in the longitudinal wavenumber of strong enough modulation is limited to $\Delta k_z\sim \alpha/\sigma_E$, which requires strong RF cavities
for significant compression of modulations (increasing the longitudinal wavenumber by an order of magnitude).

\begin{figure}[ht]
\includegraphics[width=2.5in,keepaspectratio]{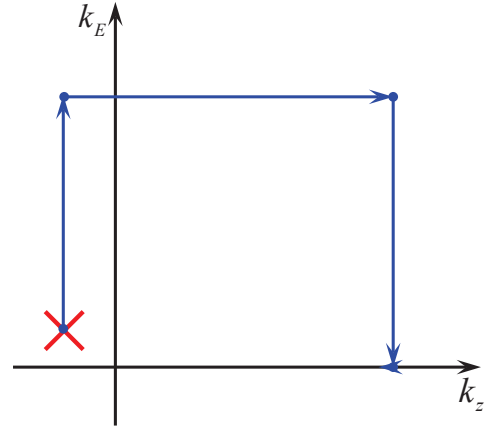}
\caption{(Color online) Schematics for the modulation transform in the CHG scheme.}
\label{fig:CHG}
\end{figure}

To overcome this limit of the CHG scheme, the second chicane was introduced in the beamline \cite{CHG}. The first 
chicane increases $k_E$ component of the modulation so that the following cavity would result in larger
shift of the longitudinal wavenumber $k_z$ as follows from Eq.~(\ref{cavity}). Once the longitudinal wavenumber of modulation is increased, the second chicane is used to recover modulation as
purely longitudinal bunching. The second RF cavity can be optionally added to the beamline to eliminate the energy chirp from the beam. This cavity does not change the longitudinal wavenumber of modulation
($k_E=0$ for longitudinal bunching) and therefore, does not affect beam bunching once it is generated. The schematics of the described scheme is presented in Fig.~\ref{fig:CHG}. The sequence
of the RF cavities and chicanes duplicates the design described in Ref.~\cite{CHG}. The final bunching is recovered from the $n=-1$ harmonic of the laser modulation in order to utilize chicanes with
positive energy dispersion.

The scheme showed in Fig.~\ref{fig:CHG} is designed to transform initial modulation having a given wavenumber $\kk^0=(-\overline{k_\varphi}, (k_E^{(n)})_{\rm max})^T$ into longitudinal bunching having the
wavevector $\kk=(M\overline{k_\varphi},0)^T$, where $M$ is the modulation compression factor. Such a transform can be accomplished with a linear beamline consisting of two chicanes and the RF cavity
accelerating the beam at X-point
\be
\kk=\left(R_{\rm chicane 2}R_{\rm cavity 1}R_{\rm chicane 1}\right)^{-T}\kk^0.
\ee
Two equations relating components of the modulation wavevector provide relationships between three parameters of the beamline (dispersions of two chicanes and the RF cavity strength). Assuming that the
amplitude of energy modulation significantly exceeds the rms beam energy, {\it i.e.} $(k_E^{(n)})_{\rm max}=\mu_{11}/\delta E$, one finds
\bea
&&\xi_1=\frac{M+1}{-\alpha_1}-\frac{\mu_{11}}{\delta E},\\
&&\xi_2=\frac{M+1}{M}\frac{1}{-\alpha_1}.
\eea
Note that the required cavity strength is negative to ensure positive chicane dispersion. Also note that the strengths of both chicanes are inversely proportional to the amplitude of the RF field. Therefore,
this scheme is a subject for parameters study to ensure that all the elements can be made linear enough.


The bunching bandwidth can be estimated using invariant (\ref{bandwidth}). The imposed modulation is mostly longitudinal, $\overline{k_\varphi}^2\sigma_z^2\gg\mu_{11}^2\sigma_E^2/(\delta E)^2$ so the
compression can also be considered as mostly longitudinal. Then the relative bandwidth of the modulation is preserved along the linear beamline and one can estimate CHG bandwidth as
\be
\left(\frac{\overline{\delta k^2}}{\overline{k}^2}\right)_{\rm CHG}^{(M)} \approx \frac{\overline{\delta k_\varphi^2}}{\overline{k_\varphi}^2}.
\ee
The relative bunching bandwidth mirrors the relative bandwidth of the laser pulse having phase noise and it does not depend on the beam compression factor $M$.
Note that this scheme requires smaller laser bandwidth compared to HGHG and EEHG schemes in order to achieve the same bandwidth of the resulting bunching.

\section{Conclusions}
In this paper we developed formalism for quantitative description of modulated beams dynamics.
This formalism is based on introduction of the spectral distribution function which is equal to the Fourier transform of the
beam distribution function in the 6D phase space. The introduced spectral distribution is localized in some region of the spectral domain which makes possible to characterize it with a limited number
of parameters such as the average wavevector of modulation and the bandwidth matrix. The key advantage of the proposed formalism is based on the observation that any given spectral component of the
distribution remains as a single component under linear transforms indicating that the topology of the spectral distribution remains the same along linear beamline. As a result, the characteristic
parameters of the spectral distribution transform linearly. 

The developed formalism has two main advantages compared to alternative approaches based on trailing coordinates of each individual electron in the phase space. First, the formalism quantifies the
main parameters of modulated beams which introduces suitable metrics characterizing modulated beams. Different competing schemes can be easily compared to each other using this metrics. Second,
the developed formalism describes the change of the spectral parameters along arbitrary linear beamline.
The spectral parameters transform linearly and their change is solely described with the conventional beam transform matrix.
As a result, the developed description of modulated beams dynamics is full and self-consistent.

This description estimates the bandwidth of the resulting modulation which allows to determine whether some proposed scheme satisfies requirements for a given application. 
As an example, we considered several laser based schemes proposed for XFEL seeding and found the output bunching bandwidth if the laser has random phase noise.

Note that description of the beam in the spectral domain of the phase space is a useful concept for wide class of problems in beam physics. It can be applied for solving homogeneous linear
problems \cite{ISR_debunching} or analyzing linear instabilities.

\section{Acknowledgements}
Author is thankful to B.~E.~Carlsten for useful discussions and suggestions. This work is supported by the U.S. Department of Energy through the LANL/LDRD program.

\appendix
\section{Interpretation of modulation invariants}
\label{app:inv}

Invariants (\ref{inv}) of the modulated beam dynamics can be interpreted using the eigen-emittance concept \cite{Dragt_eig}.
According to Williamson's theorem \cite{Williamson}, any real symmetric positively defined matrix $\Sigma$ can be diagonalized
with a symplectic matrix $A$, {\it i.e.} $A^TJA=J$, through the transform
\bea
\label{diag}
&&A(s)\Sigma(s) A^T(s)=\Lambda,\;\;\;\;\;\;\;\Lambda_{ij}=0\;\;{\rm for}\;i\neq j,\\
&&\Lambda_{2n-1,2n-1}\cdot\Lambda_{2n,2n}=\epsilon_{\rm eig}^{(n)}.
\eea
In case of the rms beam matrix $\Sigma=<\zzeta\zzeta^T>$,  $\epsilon_{\rm eig}^{(n)}$ are called the ``eigen-emittances" and they remain constant along the linear beamline.
Williamson's theorem can be interpreted in two ways. The first interpretation states that any arbitrary beam matrix can be diagonalized throughout appropriate beam optics which transform matrix is equal to
$R=A$. Alternatively, matrix $A$ can be considered as canonical transformation of canonical variables $\zzeta_{\rm eig}=A\zzeta$ in which the beam matrix becomes diagonal. Williamson's Theorem states
that the transform matrix $A$ can be chosen in such a way that the diagonal terms come in pairs, {\it i.e.} $\Lambda_{2n-1,2n-1}=\Lambda_{2n,2n}$. However, that requires renormalization of the
canonical conjugate variables so that they have equal dimensions. As a result, the elements of the beam matrix lack of clear physical interpretation such as rms bunch sizes or angular spread. Alternatively,
we chose transform matrix $A$ which only decouples the eigen- phase planes but does not necessarily diagonalize the beam matrices associated with these planes
\bea
&&A(s)\Sigma(s) A^T(s)=diag\left(\Sigma^{(n)}(s)\right),\\
&&\det\left(\Sigma^{(n)}(s)\right)=\epsilon_{\rm eig}^{(n)}=inv,
\eea
where $\Sigma^{(n)}(s)$ is $2\times2$ beam matrices corresponding to $n$-th eigen- phase plane.

The concept of eigen- phase planes provides simple picture for bunch dynamics since particle motion in different eigen- phase planes is decoupled from each other.
Note that decomposition into the eigen- phase space coordinates is not unique 
and so the beam dynamics within a given eigen- phase plane depends on the choice the of eigen-coordinates. For example, if one chooses the eigen-coordinates $\zzeta_{\rm eig}=A(s)\zzeta$ so that
$A(s)=A(s_0)R(s,s_0)^{-1}$ then the beam is stationary in this rotating frame, {\it i.e.} $\partial_s \Sigma^{(n)}(s)=0$. This choice of rotating phase space may be not the simplest description. For example,
quad focusing and diffraction are easier to describe in a stationary frame.

The eigen- phase plane concept also simplifies description of modulated beams. The change of the phase space coordinates from the lab frame to the eigen-coordinates $\zzeta_{\rm eig}=A(s)\zzeta$ defines
the wavevector of modulation in this frame ${\bf k}_{\rm eig}(s)=A^{-T}(s){\bf k}(s)$. One can naturally define the projections of the wavevector onto the eigen- phase planes as 
${\bf k}_{\rm eig}^{(n)}=(({ k}_{\rm eig})_{2n-1},({ k}_{\rm eig})_{2n})^T$. These projections of the wavevector evolve independently from each other since particle dynamics within different
eigen- phase planes is not coupled. Moreover, if one chooses rotating frame such as $A(s)=A(s_0)R(s,s_0)^{-1}$ then the wavevector of modulation in this frame
does not change along the beamline. As a result,
the entire distribution function remains constant in this frame since each individual Fourier harmonic does not change. The choice of rotating frame which does not eliminate particle dynamics within
each eigen- phase plane, $\partial_s \Sigma^{(n)}(s)\neq0$, also implies evolution of the eigen-wavevector, $\partial_s {\bf k}^{(n)}(s)\neq0$. However, in this case one can find the invariants
\be
\left({\bf k}^{(n)}(s)\right)^T \Sigma^{(n)}(s){\bf k}^{(n)}(s)=inv.
\ee
These invariants can be interpreted as the rms phase of the modulation across each eigen- phase plane area.

The same invariants can be formulated in the lab frame without introducing eigen-coordinates. In this case the invariants read as
\bea
&{\bf k}^T(s)J\left(\Sigma^{-1}(s)J\right)^{2n+1}{\bf k}(s)=inv,\\
\nonumber
&n=1,2,\, ... \, dim(\Sigma)/2.&
\eea
There is infinite number of invariants but the number of functionally independent ones is equal to the dimension of space describing beam dynamics, {\it i.e.} is equal to 3 in general case. Moreover,
preservation of the rms modulation phase across each eigen- phase plane leads to preservation of the rms modulation phase across the entire phase space volume
\be
\sum\limits_{n=1}\limits^{dim(\Sigma)/2}\left({\bf k}^{(n)}\right)^T \Sigma^{(n)}{\bf k}^{(n)}={\bf k}^T\Sigma{\bf k}=inv.
\ee



\begin{thebibliography}{8}

\bibitem{FLASH}
W.~Ackermann et al.,  Nature Photon. {\bf 1}, 336 (2007).

\bibitem{LCLS}
P.~Emma et al., Nature Photon. {\bf 4}, 641 (2010).

\bibitem{Spring8}
H.~Tanaka, Proceedings of 33rd International FEL Conference (2011); T.~Tanaka and T.~Shintake, SCSS XFEL Conceptual Design Report (Riken
Harima Institute, 2005).

\bibitem{EuroXFEL}
M.~Altarelli et al.  XFEL: The European X-Ray Free-Electron Laser
Technical Design Report. Preprint DESY 2006-097 (DESY, 2006).


\bibitem{FERMI}
E.~Allaria, C.~Callegari, D.~Cocco1, W.~M.~Fawley, M.~Kiskinova, C.~Masciovecchio, and F.~Parmigiani, New J. Phys. {\bf 12}, 075002 (2010).

\bibitem{SASE}
A.~M.~Kondratenko and E.~L.~Saldin, Part. Accel. {\bf 10}, 207 (1980).

\bibitem{optical_seeding}
J.~Feldhaus, E.~L.~Saldin, J.~R.~Schneider, E.~A.~Schneidmiller, and M.~V.~Yurkov, Opt. Comm. {\bf 140}, 341 (1997).

\bibitem{HGHG}
L.~Yu, Phys. Rev. A {\bf 44}, 5178 (1991).

\bibitem{EEHG}
G.~Stupakov, Phys. Rev. Lett. {\bf 102}, 074801 (2009); D.~Xiang and G.~Stupakov, Phys. Rev. ST Accel. Beams {\bf 12}, 030702 (2009).

\bibitem{CHG}
S.~G.~Biedrona, S.~V.~Miltona, and H.~P.~Freund, Nucl. Instr. and Meth. A {\bf 475}, 401 (2001).

\bibitem{EEHG-EEX}
B.~Jiang, J.~G.~Power, R.~Lindberg, W.~Liu,~ and W.~Gai, Phys. Rev. Lett. {\bf 106}, 114801 (2011).

\bibitem{JMO}
B.~E.~Carlsten et al., J. Modern Optics {\bf 58}, 1374 (2011).

\bibitem{HGHG_exp}
L.-H.~Yu et al., Science {\bf 289}, 932 (2000).

\bibitem{EEX_exp}
Y.-E.~Sun, P.~Piot, A.~Johnson, A.~H.~Lumpkin, T.~J.~Maxwell, J.~Ruan, and R.~Thurman-Keup, Phys. Rev. Lett. {\bf 105}, 234801 (2010).

\bibitem{EEHG_exp}
D.~Xiang et. al., Phys. Rev. Lett. {\bf 105}, 114801 (2010).

\bibitem{Brookhaven_dogleg}
P.~Muggli, V.~Yakimenko, M.~Babzien, E.~Kallos, and K.~ P.~Kusche, Phys. Rev. Lett. {\bf 101}, 054801 (2008).

\bibitem{Dragt_eig}
A.~J.~Dragt, F.~Neri, and G.~Rangarajan, Phys. Rev. A {\bf 45}, 2572 (1992). 

\bibitem{EEX}
M.~Cornacchia and P.~Emma, Phys. Rev. ST Accel. Beams {\bf 5}, 084001 (2002).


\bibitem{CHG_Ratner}
D.~Ratner, A.~Chao, and Z.~Huang, Phys. Rev. ST Accel. Beams {\bf 14}, 020701 (2011).

\bibitem{ISR_debunching}
N.~A.~Yampolsky and B.~E.~Carlsten, submitted to Phys. Rev. ST Accel. Beams.

\bibitem{Williamson}
J.~Williamson, Amer. J. Math. {\bf 58}, 141 (1936).



\end{thebibliography}
\end{document}